%% file: tune.tex
\documentclass[conference]{IEEEtran}
\usepackage{amsmath}
\usepackage{amsthm}
\usepackage{amssymb}
\usepackage{bm}
\usepackage{xspace}
\usepackage{xcolor}
\usepackage{graphicx}
\usepackage{url}
\usepackage{cite}
\usepackage{framed}
\usepackage{float}
\usepackage{rotating}

\newcommand{\executeiffilenewer}[3]{%
\ifnum\pdfstrcmp{\pdffilemoddate{#1}}%
{\pdffilemoddate{#2}}>0%
{\immediate\write18{#3}}\fi%
}

\graphicspath{{images/}}

\newcounter{algocount}

\newenvironment{algorithm}[1][]{\refstepcounter{algocount}\begin{trivlist}\item \textbf{Algorithm \thealgocount.}#1\\[-0.2cm]\rule{\columnwidth}{1pt}}{\\[-0.2cm]\rule{\columnwidth}{1pt}\end{trivlist}}

\theoremstyle{plain}

\theoremstyle{definition}

\theoremstyle{plain}

\theoremstyle{definition}

\theoremstyle{remark}

\newcommand{\vecone}{\boldsymbol{1}}

\newcommand{\vecpi}{\boldsymbol{\pi}}
\newcommand{\veca}{\boldsymbol{a}}
\newcommand{\vecc}{\boldsymbol{c}}
\newcommand{\vecd}{\boldsymbol{d}}

\newcommand{\vecn}{\boldsymbol{n}}
\newcommand{\vecp}{\boldsymbol{p}}
\newcommand{\vecq}{\boldsymbol{q}}
\newcommand{\vecr}{\boldsymbol{r}}

\newcommand{\vecw}{\boldsymbol{w}}
\newcommand{\vecx}{\boldsymbol{x}}

\DeclareMathOperator{\kl}{D}

\DeclareMathOperator*{\minimize}{minimize}

\DeclareMathOperator*{\st}{subject\,to}

\title{Short Huffman Codes Producing 1\!s Half of the Time}

\IEEEoverridecommandlockouts

\author{\IEEEauthorblockN{Fabian Altenbach, Georg B\"ocherer, and Rudolf Mathar}
\IEEEauthorblockA{Institute for Theoretical Information
Technology\\
RWTH Aachen University, 52056 Aachen, Germany
\\ Email: \texttt{\{altenbach,boecherer,mathar\}@ti.rwth-aachen.de}}
\thanks{This work has been supported by the UMIC Research Center, RWTH
Aachen University.}
}

\begin{document}
\maketitle
\input{abstract}

\input{introduction}

\input{problem}

\input{approach}

\input{optimization}

\input{results}

\bibliographystyle{IEEEtran}
\normalsize
\bibliography{IEEEabrv,confs-jrnls,Dnc}

\end{document}

%% file: abstract.tex
\begin{abstract}
The design of the channel part of a digital communication system (e.g., error correction, modulation) is heavily based on the assumption that the data to be transmitted forms a fair bit stream. However, simple source encoders such as short Huffman codes generate bit streams that poorly match this assumption. As a result, the channel input distribution does not match the original design criteria. In this work, a simple method called half Huffman coding (\textsc{halfHc}) is developed. \textsc{halfHc} transforms a Huffman code into a source code whose output is more similar to a fair bit stream. This is achieved by permuting the codewords such that the frequency of 1s at the output is close to 0.5. The permutations are such that the optimality in terms of achieved compression ratio is preserved. \textsc{halfHc} is applied in a practical example, and the resulting overall system performs better than when conventional Huffman coding is used.
\end{abstract}

%% file: introduction.tex
\section{Introduction}

The key part of a digital communication system is the binary interface between source coding and channel coding \text{\cite[p. 1--3]{Gallager2008}}. The ultimate objective of source coding is to transform the data into a sequence of 0s and 1s that is indistinguishable from a fair bit stream, i.e., a sequence of independent and identically distributed (iid) equiprobable bits. Accordingly, most work on channel coding starts with the assumption that the data to be transmitted comes as a fair bit stream.

In \cite{Bocherer2011}, we have shown that the capacity achieving input probability mass function (pmf) of a channel can be approximated arbitrarily well by parsing a fair bit stream by a matcher code and by doing so, capacity achieving modulation can be established \cite{Bocherer2011a}. However, this result is heavily based on the assumption that the binary interface is a fair bit stream. Asymptotically in terms of source encoder complexity, the binary interface can indeed be turned into a fair bit stream \cite{Han2005}. In practice, however, we are often restricted to simple encoders, and the resulting binary output differs significantly from a fair bit stream. As a simple measure to evaluate how close the generated bit stream is to a fair bit stream, the frequency of generated $1$s can be considered. For a fair bit stream, this frequency should be $0.5$. One possibility to achieve this goal is to pass the source encoder output through a scrambler \cite{Ungerboeck2002}. The drawback of this approach is that the corresponding descrambler has to be known and implemented at the receiver \cite{Ungerboeck2008}.

In this work we propose an algorithm called \emph{half Huffman coding} (\textsc{halfHc}) that constructs prefix-free source codes. \textsc{halfHc} achieves the optimal compression ratio (i.e., the same as Huffman coding (\textsc{Hc})), and the frequency of $1$s of \textsc{halfHc} is closer to 0.5 than the frequency of $1$s of \textsc{Hc}. As in the case of conventional \textsc{Hc}, no additional descrambler at the receiver is required. We apply \textsc{halfHc} to the shaping problem in \cite{Bocherer2011c}. We show that the resulting channel input pmf is closer to the one predicted by the fair bit stream assumption than the channel input pmf resulting from conventional \textsc{Hc}. A complete implementation of \textsc{halfHc} in Matlab can be found at our website \cite{website:halfhc}.

The remainder of this paper is organized as follows. In Section~\ref{sec:problem}, we state the main problem. The idea of our solution is formulated in Section~\ref{sec:idea}. We then formulate in Section~\ref{sec:half} our idea as a mathematical optimization problem, propose methods for solving it, and formulate our new algorithm \textsc{halfHc}. Finally, in Section~\ref{sec:results} we apply \textsc{halfHc} to a practical example.

%% file: problem.tex
\section{Problem Statement}
\label{sec:problem}
\subsection{Motivating example}
\begin{figure}
\footnotesize
\centering
\def\svgwidth{1.0\columnwidth}
\executeiffilenewer{images/system.svg}{images/system.pdf}%
{inkscape -z -D --file=images/system.svg %
--export-pdf=images/system.pdf --export-latex}%
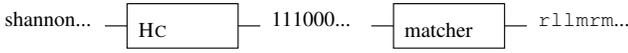%

\caption{We first compress the text to a binary sequence by \textsc{Hc} and then match the binary sequence to the design criteria by a matcher code. At the output of the matcher, a sequence of the symbols \texttt{r,l,m} is generated.}
\label{fig:system}
\end{figure}

In \cite{Bocherer2011c}, we consider a channel with three input symbols \texttt{r,l,m} with the associated costs
\begin{align}
\vecw&=(0.18,\;0.18,\;0.31)^T,
\end{align}
where $(\cdot)^T$ is the transpose. The channel input pmf $\vecp$ is subject to the cost constraint
\begin{align}
\vecw^T\vecp\leq S=0.2063.
\end{align}
The optimal channel input probability mass function (pmf) was calculated as
\begin{align}
\vecp^* &= (0.3988,\; 0.3988,\;  0.2023)^T.
\end{align} 
Note that $\vecp^*$ fulfills the cost constraint with equality. The class of pmfs that can be generated at the channel input by parsing a fair bit stream by a matcher code are the \emph{dyadic pmfs} \cite{Bocherer2011}. A pmf is dyadic if each entry $p_i$ can be written in the form
\begin{align}
p_i = 2^{-\ell},\quad\ell\in\mathbb{N}
\end{align}
i.e., each entry is a positive integer power of $1/2$. The optimization problem in \cite{Bocherer2011} of finding an optimal matcher code can be stated as
\begin{align}
\begin{split}
\minimize\quad&\kl(\vecd\Vert\vecp^*)\\
\st\quad&\vecw^T\vecd\leq S\\
&\vecd\text{ is a dyadic pmf},
\end{split}\label{eq:matching}
\end{align}
where $\kl(\cdot\Vert\cdot)$ is the Kullback-Leibler (KL) distance, which is defined as 
\begin{align}
 \kl(\vecp\Vert\vecq) & = \sum_i p_i \log \frac{p_i}{q_i},
\end{align}
with $\log$ denoting the natural logarithm. In \cite{Bocherer2011c}, the algorithm \emph{cost constrained geometric Huffman coding} (\textsc{ccGhc}) is developed, and it is shown that for a fair bit stream at the binary interface, the resulting matcher code optimally solves Problem~\eqref{eq:matching}. The solution $\vecd$ achieves
\begin{align}
\kl(\vecd\Vert\vecp^*)&=0.0048392\\
\vecw^T\vecd &= 0.20607\leq S.
\end{align}
However, the bit stream at the binary interface is generated by compressing the English text \cite{Quotes} by a simple \textsc{Hc}, and this bit stream is then parsed by the matcher code, see Fig.~\ref{fig:system}. By \textsc{Hc} we refer to Huffman coding as implemented by \texttt{huffmandict.m} in the Matlab Communications System Toolbox.
The effective pmf, the effective KL-distance, and the effective cost at the output of the matcher are respectively given by
\begin{align}
\vecd_\textsc{Hc}&=(0.40663, \;0.35761, \;0.23576)^T \label{eq:obs1}\\
\kl(\vecd_\textsc{Hc}\Vert\vecp^*)&=0.0050066 \label{eq:obs2}\\
\vecw^T\vecd_\textsc{Hc} &=  0.21065 > S \label{eq:obs3}
\end{align}
i.e., what is actually observed significantly differs from what we would expect under the fair bit stream assumption.

\subsection{Approach}
While observations~\eqref{eq:obs1}--\eqref{eq:obs3} may occur since the effectively observed values are a realization of a random process, which can always diverge from the expected values, the reason can also be that the assumption of a fair bit stream is false. To evaluate the quality of the binary interface, we calculate the effective relative number of 1s. For a fair bit stream, this number should be close to $q=0.5$. However, the effective frequency that we observe after applying \textsc{Hc} to the English text \cite{Quotes} is 
\begin{align}
q_\textsc{Hc} = 0.45821.
\end{align}
A maximum likelihood estimator (MLE) of $q$ yields for the bit stream at the output of \textsc{Hc} the $95\%$ confidence interval $(0.4464,\;0.47006)$. Since the corresponding value of $q=0.5$ is not contained in this interval, the MLE rejects the hypothesis of having a fair bit stream after applying \textsc{Hc}. Our approach is therefore to look for optimal prefix-free source codes that yield a $q$ close to $0.5$.

%% file: images/system.pdf_tex

\begingroup
  \makeatletter
  \providecommand\color[2][]{%
    \errmessage{(Inkscape) Color is used for the text in Inkscape, but the package 'color.sty' is not loaded}
    \renewcommand\color[2][]{}%
  }
  \providecommand\transparent[1]{%
    \errmessage{(Inkscape) Transparency is used (non-zero) for the text in Inkscape, but the package 'transparent.sty' is not loaded}
    \renewcommand\transparent[1]{}%
  }
  \providecommand\rotatebox[2]{#2}
  \ifx\svgwidth\undefined
    \setlength{\unitlength}{480pt}
  \else
    \setlength{\unitlength}{\svgwidth}
  \fi
  \global\let\svgwidth\undefined
  \makeatother
  \begin{picture}(1,0.06833333)%
    \put(0,0){\includegraphics[width=\unitlength]{system.pdf}}%
    \put(0,0.03418803){\color[rgb]{0,0,0}\makebox(0,0)[lb]{\smash{shannon...}}}%
    \put(0.2,0.01752136){\color[rgb]{0,0,0}\makebox(0,0)[lb]{\smash{\textsc{Hc}}}}%
    \put(0.4,0.03418803){\color[rgb]{0,0,0}\makebox(0,0)[lb]{\smash{111000...}}}%
    \put(0.6,0.01752136){\color[rgb]{0,0,0}\makebox(0,0)[lb]{\smash{matcher}}}%
    \put(0.8,0.03418803){\color[rgb]{0,0,0}\makebox(0,0)[lb]{\smash{\texttt{rllmrm}...}}}%
  \end{picture}%
\endgroup

%% file: approach.tex
\section{Main Idea}
\label{sec:idea}

\begin{table}
\caption{\textsc{Hc} for the English text \cite{Quotes}}
\label{tab:codes}
\centering
\begin{tabular}{rllllc}
&$i$&$p_i$&$\vecc_i$&$l(i)$&$j$\\\hline
\input{m/code.txt}
\end{tabular}
\end{table}

As stated in \cite[Sec. 5.8]{Cover2006}, optimal prefix-free source codes are not unique and for a given pmf, \textsc{Hc} constructs \emph{one} optimal code. All other optimal codes can be obtained by applying appropriate permutations of the code generated by \textsc{Hc}. In this section we will show how the frequency of 1s can be influenced by specific permutations.

\subsection{Frequency of 1s}

Denote the number of symbols of the considered source by $n$. Without loss of generality, we assume that the pmf $\vecp$ describing the source is ordered with decreasing probabilities, i.e,
\begin{align}
p_1\geq p_2\geq\dotsb\geq p_n.
\end{align}
Denote by $\mathcal{C}$ the ordered set of codewords generated by \textsc{Hc}, i.e.,
\begin{align}
\mathcal{C} = (\vecc_1,\dotsc,\vecc_n).
\end{align}
Denote by $\vecone^T\vecc_i$ the number of 1s in the $i$th codeword. We write the length of codeword $\vecc_i$ as $l(i)$. Denote by $m$ the number of distinct codeword lengths and order and index the set of distinct codeword lengths $\{\ell_j\}_{j=1}^m$. For the English text \cite{Quotes} we have $i,p_i,\vecc_i,l(i)$ and $j$ displayed in Table~\ref{tab:codes}. Denote the pmf of the codeword lengths by $\vecr$, i.e.,
\begin{align}
r_j = \sum_{i:l(i)=\ell_j} p_i,\qquad j=1,\dotsc,m.
\end{align}
Denote by $p_{i|j}$ the probability that symbol $i$ is generated given that the codeword length is $\ell_j$, i.e., if $p_{i|j}>0$, then
\begin{align}
 p_{i|j}=\frac{p_i}{r_j}.
\end{align}
Denote by $N_j$ the expected number of 1s conditioned on codeword length $\ell_j$, i.e.,
\begin{align}
N_j = \sum_{i:l(i)=\ell_j}  p_{i|j}\vecone^T\vecc_i.
\end{align}
Thus the expected frequency $q$ of 1s given by the expected number of 1s $N$ per average length $L$ can be written as
\begin{align}
q & = \frac{N}{L}\\
&=\frac{\sum\limits_{j=1}^m r_j N_j}{\sum\limits_{j=1}^m\sum\limits_{i:l(i)=\ell_j} p_{i|j}r_j \ell_j}.
\end{align}

\subsection{Permutations of $\mathcal{C}$}

Any permutation of codewords of the same length again gives an optimal code. However, while the achieved compression rate is invariant under these permutation, the mean number of 1s is not. This is because in general, there are codewords of the same length that occur with different probabilities. A naive approach consists of searching through all permutations, and choose the one with the mean number of 1s being closest to $0,5$. However, the number of such permutations becomes very fast prohibitively large. For instance, for the code displayed in Table~\ref{tab:codes}, this number is approximately
\begin{align}
\label{eq:perm_huge}
2!7!7!5!2!4!\approx3\cdot10^{11}.
\end{align}
Consider now the set of codewords $\mathcal{C}_j$ of codewords of length $\ell_j$. We consider two special permutations of $\mathcal{C}_j$. First, the one that \emph{maximizes} the expected number of 1s and second, the one that \emph{minimizes} the expected number of 1s. The first is obtained by ordering the codewords in $\mathcal{C}_j$ with decreasing number of 1s, and the second one is obtained by ordering the codewords in $\mathcal{C}_j$ with increasing number of 1s. Denote the corresponding permutations by $\pi_j^+$ and $\pi_j^-$, respectively. For example, for the code in Table~\ref{tab:codes}, the corresponding permutations for codeword length $\ell_1 = 3$ are 
\begin{align}
 \pi_1^+: 1 &\mapsto 2, \hspace{0.25cm}2 \mapsto 1 \\ \pi_1^-: 1 &\mapsto 1, \hspace{0.25cm}2 \mapsto 2.
\end{align}
The maximum expected number $N^+_j$ of 1s in a codeword, conditioned on the codeword length $\ell_j$ is given by
\begin{align}
N_j^+=\sum_{i:l(i)=\ell_j} p_{i|j}\vecone^T\vecc_{\pi_j^+(i)}\label{eq:permPlus}
\end{align}
and the minimum expected number of 1s $N_j^-$ is given by
\begin{align}
N_j^-=\sum_{i:l(i)=\ell_j} p_{i|j}\vecone^T\vecc_{\pi_j^-(i)}.\label{eq:permMinus}
\end{align}
For example, in case of codeword length 3 (corresponding to $\ell_1$), we can see from Table \ref{tab:codes} that
\begin{align*}
N^+_1 = \frac{p_1}{p_1+p_2}\cdot 2 + \frac{p_2}{p_1+p_2}\cdot 0 \approx 1.276 \\
N^-_1 = \frac{p_1}{p_1+p_2}\cdot 0 + \frac{p_2}{p_1+p_2}\cdot 2 \approx 0.724
\end{align*}
The idea is now to choose for each codeword length $\ell_j$ between $N_j^+$ and $N_j^-$, with the objective to get the overall expected number of 1s divided by the expected codeword length as close to $q=0.5$ as possible. This means, we would like to solve
\begin{align}
\label{eq:gaul}
q& = \frac{\sum\limits_{j=1}^m r_jN_j^{ z_j}}{L}\overset{!}{\approx} 0.5
\end{align}
where each $z_j$ takes values in $\{+,-\}$. The corresponding optimization problem is discussed in the next section.

%% file: m/code.txt
\_  & 1 & 0.1699 & 000 & 3 & 1\\e  & 2 & 0.0964 & 110 & 3 & 1\\t  & 3 & 0.0777 & 0010 & 4 & 2\\a  & 4 & 0.0717 & 0100 & 4 & 2\\i  & 5 & 0.0663 & 0101 & 4 & 2\\o  & 6 & 0.0645 & 0111 & 4 & 2\\n  & 7 & 0.0614 & 1000 & 4 & 2\\r  & 8 & 0.0530 & 1010 & 4 & 2\\s  & 9 & 0.0500 & 1110 & 4 & 2\\h  & 10 & 0.0373 & 00110 & 5 & 3\\c  & 11 & 0.0325 & 01101 & 5 & 3\\l  & 12 & 0.0325 & 01100 & 5 & 3\\m  & 13 & 0.0277 & 10110 & 5 & 3\\u  & 14 & 0.0277 & 10011 & 5 & 3\\d  & 15 & 0.0235 & 11110 & 5 & 3\\f  & 16 & 0.0199 & 11111 & 5 & 3\\g  & 17 & 0.0181 & 001110 & 6 & 4\\y  & 18 & 0.0145 & 100100 & 6 & 4\\p  & 19 & 0.0139 & 100101 & 6 & 4\\b  & 20 & 0.0133 & 101110 & 6 & 4\\w  & 21 & 0.0114 & 101111 & 6 & 4\\v  & 22 & 0.0054 & 00111100 & 8 & 5\\k  & 23 & 0.0042 & 00111101 & 8 & 5\\x  & 24 & 0.0024 & 001111100 & 9 & 6\\q  & 25 & 0.0018 & 001111110 & 9 & 6\\z  & 26 & 0.0018 & 001111101 & 9 & 6\\j  & 27 & 0.0012 & 001111111 & 9 & 6

%% file: optimization.tex
\section{Half Huffman Coding}
\label{sec:half}

We will express goal~\eqref{eq:gaul} in terms of a mathematical optimization problem. For each index $j$ corresponding to a specific codeword length $\ell_j$, we can either take the maximizing permutation $\pi_j^+$ or the minimizing permutation $\pi_j^-$. We introduce a binary vector $\vecx \in \{0,1\}^m$, which serves as a selection variable between both permutations. By using~\eqref{eq:gaul}, the expected frequency of 1s can be expressed in terms of $x_j$ as
\begin{align}
q& = \frac{\sum\limits_{j=1}^m r_jN_j^{z_j}}{L}\\
& = \frac{1}{L} \sum_{j=1}^m r_j \left[x_j(N_j^- - N_j^+) + N_j^+ \right].
\end{align}
For $x_j=0$, we choose $N_j^+$, and for $x_j = 1$, we choose $N_j^-$. We measure the quality of our selection $\vecx$ by the absolute deviation between $q$ and $0.5$. Hence, in vector notation, the objective we want to minimize has the form
\begin{align}
|q - 0.5| & =  \left|\frac{1}{L} \left[ \vecr\circ(\vecn_- - \vecn_+)\right]^T \vecx + \frac{1}{L} \vecn_+^T\vecr - 0.5\right|,
\end{align}
where $\circ$ denotes the elementwise Hadamard vector product, and $\vecn_-$ and $\vecn_+$ are defined as
\begin{align}
 \vecn_- &= (N_1^-, \ldots, N_m^-)^T\\
\vecn_+ &= (N_1^+, \ldots, N_m^+)^T.
\end{align}
For notational convenience, we further substitute 
\begin{align}              
\veca &= \frac{1}{L} \left[ \vecr\circ(\vecn_- - \vecn_+)\right]\\
b &=  \frac{1}{L} \vecn_+^T\vecr - 0.5.
\end{align}
We can now state the optimization problem
\begin{equation}
\label{eq:opt_problem_abs}
\begin{array}{ll} 
\underset{\vecx}{\text{minimize}} & |\veca^T\vecx + b | \\
\text{subject to} & x_j \in \{0,1\}, \hspace{0.5cm} j = 1,\ldots,m.
\end{array}
\end{equation}
Introducing the epigraph variable $t\in\mathbb{R}_{+}$ \cite{Boyd2004}, we can directly see that the problem (\ref{eq:opt_problem_abs}) is equivalent to
\begin{equation}
\label{eq:opt_problem_MILP}
\begin{array}{ll} 
\underset{\vecx,t}{\text{minimize}} & t  \\
\text{subject to} & -t \leq \veca^T\vecx + b \leq t \\ &  x_j \in \{0,1\}, \hspace{0.5cm} j = 1,\ldots,m.
\end{array}
\end{equation}
This problem is a mixed integer linear program (MILP) in its canonical form \cite{Wolsey1999}, and is generally hard to solve. Since we are focusing on short-length Huffman codes, the number of different codeword lengths $m$ will not be too large. As discussed below, we can use standard methods for solving this problem globally. 

\subsection{Optimization methods}

\subsubsection{Naive exhaustive search}
\label{subsec:exhaustive}
In order to find the global solution to problem (\ref{eq:opt_problem_abs}), we can simply try all possible vectors $\vecx\in\{0,1\}^m$. In our example from Table~\ref{tab:codes}, $m=6$, that is, we have to choose between $2^6=64$ possible vectors $\vecx$. Thus, in our example we have overcome the huge number of possible permutations~\eqref{eq:perm_huge}, but in general, we still might be constrained by the combinatorial nature of problem (\ref{eq:opt_problem_MILP}), since the complexity of exhaustive search grows exponentially in the number of distinct codeword lengths $m$. This problem can be overcome by considering smarter search algorithms, as discussed next.

\subsubsection{Combinatorial feasibility method via bisection}
\label{subsec:feasibility}
We can also exploit some structure in the MILP formulation (\ref{eq:opt_problem_MILP}) by using a bisection method \cite[p.~146]{Boyd2004}. Suppose we set the epigraph variable to a fixed value $t$. We can now try to find a feasible solution to the remaining combinatorial feasibility problem
\begin{equation}
\label{eq:opt_problem_feas}
\begin{array}{ll} 
\text{find} & \vecx  \\
\text{subject to} & -t\leq \veca^T\vecx + b \leq t \\ &  x_j \in \{0,1\}, \hspace{0.5cm} j = 1,\ldots,m.
\end{array}
\end{equation}
There are two possible cases that can occur: 
\begin{enumerate}
\item If we find a feasible solution, the particular choice of $t$ is greater or equal than the smallest possible value. Hence the value of $t$ can be further decreased.
\item When there is no feasible solution, the choice of $t$ was too small and we have to increase it.
\end{enumerate}
After checking both cases we can solve the feasibility problem with an updated version of $t$, and repeat until convergence. This approach is summarized in Algorithm 1.
\subsubsection{Specific branch and bound method}
\label{subsec:bb}
Formulation \eqref{eq:opt_problem_abs} falls into the class of problems discussed in \cite[Sec.~2]{Boyd2007}. Thus, if none of the methods proposed in Subsection~\ref{subsec:exhaustive} and Subsection~\ref{subsec:feasibility} can solve the problem in acceptable time, a specific branch and bound solver for problem \eqref{eq:opt_problem_abs} can easily be implemented and still finds the optimal solution in hopefully reasonable time.
\begin{algorithm}\
\\
\textbf{set} $l,u$, tolerance $\epsilon > 0$\\
\textbf{repeat}\\
\indent 1. $t := (l+u)/2$\\
\indent 2. Solve the combinatorial feasibility problem (\ref{eq:opt_problem_feas})\\
\indent 3. \textbf{if} (\ref{eq:opt_problem_feas}) is feasible\\
\indent \indent \indent decrease $u := t$; \\
\indent \indent \textbf{else}\\
\indent \indent \indent increase $l := t$; \\
\textbf{until} $u-l\leq\epsilon$\\
return $\vecx$
\end{algorithm}
\subsection{Half Huffman coding}
We can now state \textsc{halfHc}, see Algorithm~\ref{alg:halfhc} for a summary. In detail, we are given a pmf with entries sorted in descending order. First, we calculate the conventional Huffman code $\mathcal{C} = \textsc{Hc}(\vecp)$. Then, for each codeword length $\ell_j$, we determine the maximum and minimum permutations $\pi_j^+$ and $\pi_j^-$, respectively. We use these permutations to calculate the vectors $\vecn_+,\vecn_-$ of maximum and minimum expected numbers of 1s. For the resulting vectors, we solve Problem~\eqref{eq:opt_problem_abs} by any method from IV.A in order to find an optimal selection vector $\vecx$. The selection vector now determines which permutation has to be applied for each codeword length $\ell_j$, i.e.,
\begin{align}
\pi_j=\begin{cases}
\pi_j^+,&\text{if }x_j = 0\\          
\pi_j^-,&\text{if }x_j = 1      
\end{cases},
\qquad j=1,\dotsc,m.
\label{eq:permSel}
\end{align}
Finally, the resulting permutation $\vecpi=(\pi_1,\dotsc,\pi_m)$ is applied to get the final code, i.e., $\textsc{halfHc}(\vecp)=\pi(\mathcal{C})$. A complete implementation of \textsc{halfHc} in Matlab can be found at our website \cite{website:halfhc}.

\begin{algorithm}[ (\textsc{halfHc})]\
\label{alg:halfhc}
$p_1\geq \dotsb \geq p_n$\\
1. $\mathcal{C}=\textsc{Hc}(\vecp)$\\
2. find $\vecn_+,\vecn_-$ via \eqref{eq:permPlus}, \eqref{eq:permMinus}\\
3. $\vecx$ = solution of \eqref{eq:opt_problem_abs} via any method from IV.A\\
4	. $\vecpi=(\pi_1,\dotsc,\pi_m)$ according to \eqref{eq:permSel}\\
return $\vecpi(\mathcal{C})$
\end{algorithm}

%% file: results.tex
\section{Numerical Results}
\label{sec:results}
\begin{table}
\caption{\textsc{halfHc} for the English text \cite{Quotes}}
\label{tab:half}
\centering
\begin{tabular}{rlllll}
&$i$&$p_i$&$\vecc_i$&$j$&$\ell_j$\\\hline
\input{m/half_code.txt}
\end{tabular}
\end{table}
We apply \textsc{halfHc} to the English text \cite{Quotes}. We execute \textsc{halfHc} twice. First, we use in step 3 exhaustive search~\ref{subsec:exhaustive}. Second, we use the combinatorial feasibility method~\ref{subsec:feasibility}. Both methods find the same selection vector $\vecx$, which is given by
\begin{align}
 \vecx & = (1, \; 0,\;0,\;0,\;0,\;0)^T.
\end{align}

The generated code is displayed in Table~\ref{tab:half}. As can be seen, for $\ell_1 = 3$, the codewords are sorted with decreasing number of 1s, while for remaining codeword lengths, the codewords with increasing number of 1s. Notice the differences to the code obtained by conventional \textsc{Hc} as displayed in Table~\ref{tab:codes}. The resulting effective frequency of 1s of \textsc{halfHc} is
\begin{align}
q_\textsc{halfHc}=0.49985.
\end{align}
This is much closer to $0.5$ than the value $0.45821$ that resulted from conventional $\textsc{Hc}$. Thus, $\textsc{halfHc}$ achieved the first objective given in~\eqref{eq:gaul}, namely to get the frequency of 1s closer to $0.5$.

Let's consider now if this has the desired effect on the effective distributions that are generated by a matcher code. For the English text \cite{Quotes}, the resulting effective pmf $\vecd_\mathrm{eff}$ is
\begin{align}
\vecd_\textsc{halfHc}=(0.38627,\;0.41107,\;0.20266)^T,\label{eq:pmfhalfhc}
\end{align}
The resulting KL-distance and average cost are respectively
\ \!\!\!\!\!\!\!\!given by
\begin{align}
\kl(\vecd_\textsc{halfHc}\Vert\vecp^*)&=0.00048629\\
\vecw^T\vecd &=   0.20635.
\end{align}
Compared to \textsc{Hc}, the KL-distance is reduced. Thus, by using \textsc{halfHc} instead of conventional \textsc{Hc}, the effective output of a matcher code is closer to the output expected under the fair bit stream assumption. The effective cost of \textsc{Hc} exceeds the cost constraint by $2.11\%$, \textsc{halfHc} exceeds the cost constraint $S=0.2063$ by only $0.02\%$. Although both \textsc{Hc} and \textsc{halfHc} formally violate the cost constraint, the value achieved by \textsc{halfHc} was adopted as a practical solution by the collaborating architects in \cite{Bocherer2011c}, who originally formulated the cost constraint $S$. We can conclude that our approach of minimizing $|q-0.5|$ leads to the desired result.
\begin{figure}[t]
\includegraphics[width=\columnwidth]{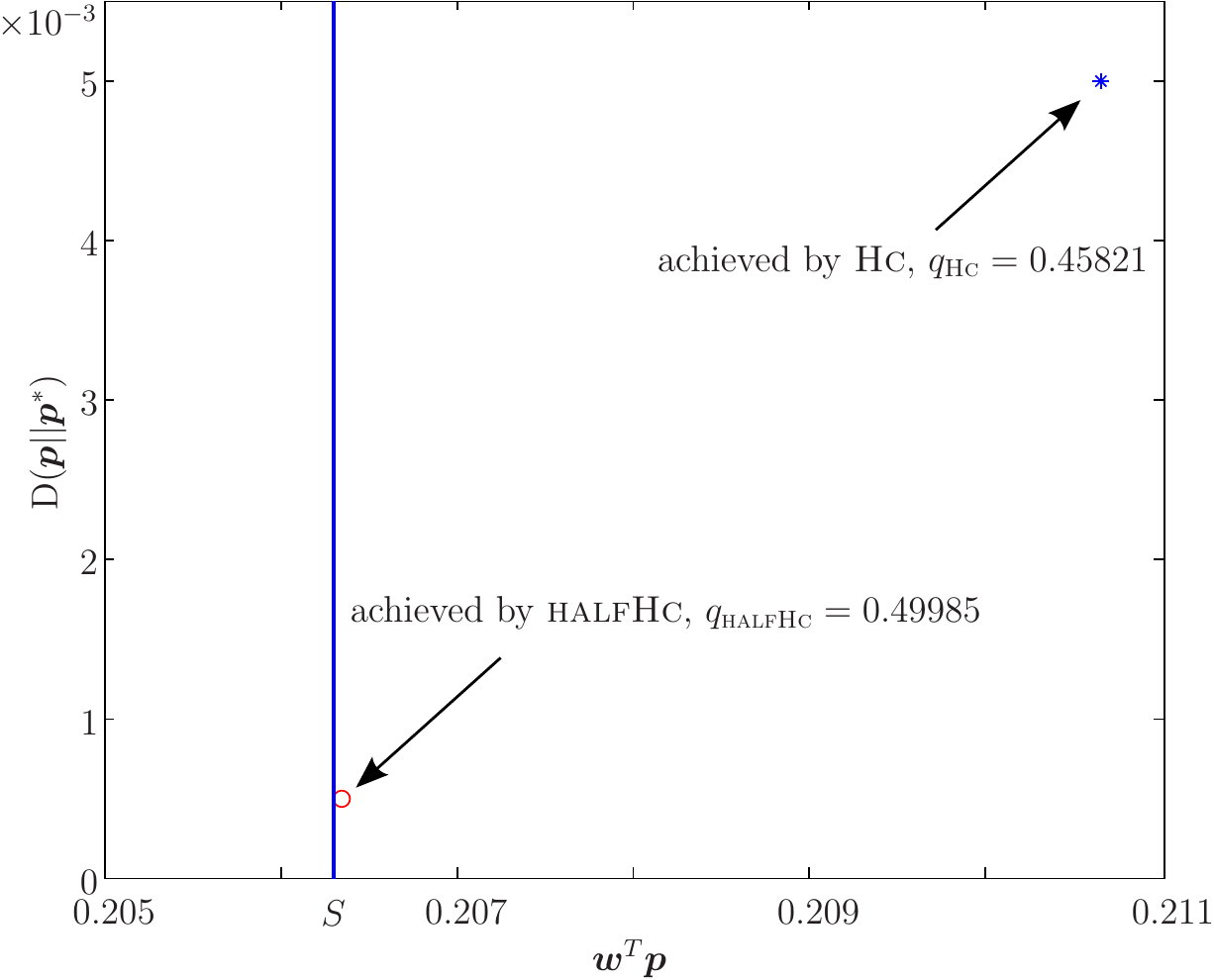}
\caption{Comparison between \textsc{Hc} and \textsc{halfHc} for the English text \cite{Quotes}. The horizontal axis corresponds to the average cost $\vecw^T\vecp$ while the vertical axis corresponds to the KL-distance $\kl(\vecp\Vert\vecp^*)$. For $\textsc{Hc}$, $\vecp=\vecd_\textsc{Hc}$, see \eqref{eq:obs1}, and for \textsc{halfHc}, $\vecp=\vecd_\textsc{halfHc}$, see \eqref{eq:pmfhalfhc}. The blue line marks the average cost constraint $S=0.2063$ of the original design problem \eqref{eq:matching}.}
\label{fig:results}
\end{figure}

%% file: m/half_code.txt
\_  & 1 & 0.1699 & 000 & 3 & 1\\e  & 2 & 0.0964 & 110 & 3 & 1\\t  & 3 & 0.0777 & 1110 & 4 & 2\\a  & 4 & 0.0717 & 0111 & 4 & 2\\i  & 5 & 0.0663 & 1010 & 4 & 2\\o  & 6 & 0.0645 & 0101 & 4 & 2\\n  & 7 & 0.0614 & 1000 & 4 & 2\\r  & 8 & 0.0530 & 0100 & 4 & 2\\s  & 9 & 0.0500 & 0010 & 4 & 2\\h  & 10 & 0.0373 & 11111 & 5 & 3\\c  & 11 & 0.0325 & 10011 & 5 & 3\\l  & 12 & 0.0325 & 11110 & 5 & 3\\m  & 13 & 0.0277 & 01101 & 5 & 3\\u  & 14 & 0.0277 & 10110 & 5 & 3\\d  & 15 & 0.0235 & 01100 & 5 & 3\\f  & 16 & 0.0199 & 00110 & 5 & 3\\g  & 17 & 0.0181 & 101111 & 6 & 4\\y  & 18 & 0.0145 & 101110 & 6 & 4\\p  & 19 & 0.0139 & 100101 & 6 & 4\\b  & 20 & 0.0133 & 001110 & 6 & 4\\w  & 21 & 0.0114 & 100100 & 6 & 4\\v  & 22 & 0.0054 & 00111101 & 8 & 5\\k  & 23 & 0.0042 & 00111100 & 8 & 5\\x  & 24 & 0.0024 & 001111111 & 9 & 6\\q  & 25 & 0.0018 & 001111110 & 9 & 6\\z  & 26 & 0.0018 & 001111101 & 9 & 6\\j  & 27 & 0.0012 & 001111100 & 9 & 6